# Emotional Responses to Auditory Hierarchical Structures is Shaped by Bodily Sensations and Listeners' Sensory Traits

Maiko Minatoya, Tatsuya Daikoku, and Yasuo Kuniyoshi
Graduate School of Information Science and Technology, The University of Tokyo, Tokyo, Japan

**Author Note**

We thank Hanna Ringer for her valuable advice on writing the manuscript.

Correspondence concerning this article should be addressed to Maiko Minatoya, Graduate School of Information Science and Technology, The University of Tokyo, Tokyo, Japan. Email: mminatoya@g.ecc.u-tokyo.ac.jp

## Competing Interests

The authors declare no competing financial interests.

## Author Contributions

M.M and T.D. conceived the experimental paradigm and method of data analysis. M.M. collected and analyzed the data and drafted the manuscript and figures. M.M. and T.D. edited and finalized the manuscript. All authors reviewed and approved the final version of the manuscript. T.D. and Y.K. supervised the project.

## Acknowledgements

This research was supported by JSPS-KAKENHI (24K22302, 24H01539, 24H00898). The funding sources had no role in the decision to publish or prepare the manuscript.

## Data Availability

The complete raw dataset and analysis results for this study can be found at https://osf.io/tpqnj/. Other data and results of statistical analysis are shown in the supplementary materials.

# Abstract

Emotional responses to auditory stimuli are a common part of everyday life. However, for some individuals, these responses can be distressing enough to interfere with daily functioning. Despite their prevalence, the mechanisms underlying auditory-induced emotion remain only partially understood. Prior research has identified contributing factors such as auditory features, listener traits, and bodily sensations. However, most studies have focused on acoustic features, leaving the role of syntactic structure largely unexplored. This study specifically investigates how hierarchical syntactic structures influence emotional experience, in conjunction with listener traits and bodily sensations. An online experiment was conducted with 715 participants, who listened to 26 sound sequences varying systematically in hierarchical syntactic complexity. Sequences were generated by combining three types of local pitch movement with three types of global pitch movement in ascending and descending pitch directions, resulting in nine complexity levels. Participants rated the valence and arousal of each sequence and indicated any bodily sensations on a body map. Measures of sensory processing patterns were also collected. Results showed that emotional valence was associated with the complex interplay of moderate syntactic complexity ("not too simple, not too complex"), sensory sensitivity, and upper torso sensations. These findings expand existing research by identifying syntactic features that shape auditory-induced emotional experience and highlight the link between bodily sensation and emotional response. They also suggest potential applications for incorporating syntactic design into therapeutic approaches to emotion regulation.

# 1. Introduction

How do emotions arise when we hear sounds? Feeling emotions as a response to sounds and sound sequences is a common part of our daily lives. We may feel scared by loud thunder, soothed by birds' chirping, or energized by rock music. Yet, how sound sequences evoke specific emotional experiences is still not completely clear.

Although the mechanism of auditory-induced emotion is not clear, such emotions have enough effect to become obstacles in daily living. Many individuals with sensory processing disorders (SPDs) suffer from negative emotional and behavioral responses induced by sensory input (*Passarello, 2022*). SPDs are linked to autism spectrum disorder (ASD), and it is known that many individuals with ASD show averse emotional and behavioral responses to auditory stimuli (*Chang, 2012*; *Fernández-Andrés, 2015*). Thus, understanding the mechanism of auditory-induced emotion is not only a theoretically, but also a clinically important question.

Various frameworks have been proposed to explain the mechanism of emotional experience induced by auditory stimuli. One of the most widely accepted frameworks was proposed by Scherer & Zentner (*2001*) (*Lennie, 2022*). We based our study on Scherer & Zentners' framework, which proposes that experienced(felt) emotion is a multiplicative function consisting of several factors: structure features (acoustic and syntactic features of the auditory stimuli), listener features (the listener's musical experience, stable disposition such as personality, and transient state such as mood), contextual features (the listening environment and social/cultural context), and performance features (the way in which the audio is played by the performer) (Figure 1). These four factors are thought to activate emotion in the brain directly through the central routes (central nervous system) and indirectly through the peripheral routes (somatic and autonomic nervous systems). Emotion induction through the peripheral route is known as peripheral feedback. Peripheral feedback was originally introduced in the James-Lange theory, which proposed that physiological arousal in the body (such as heart rate change) can affect the subjective feeling of an emotion (*James, 1884*).

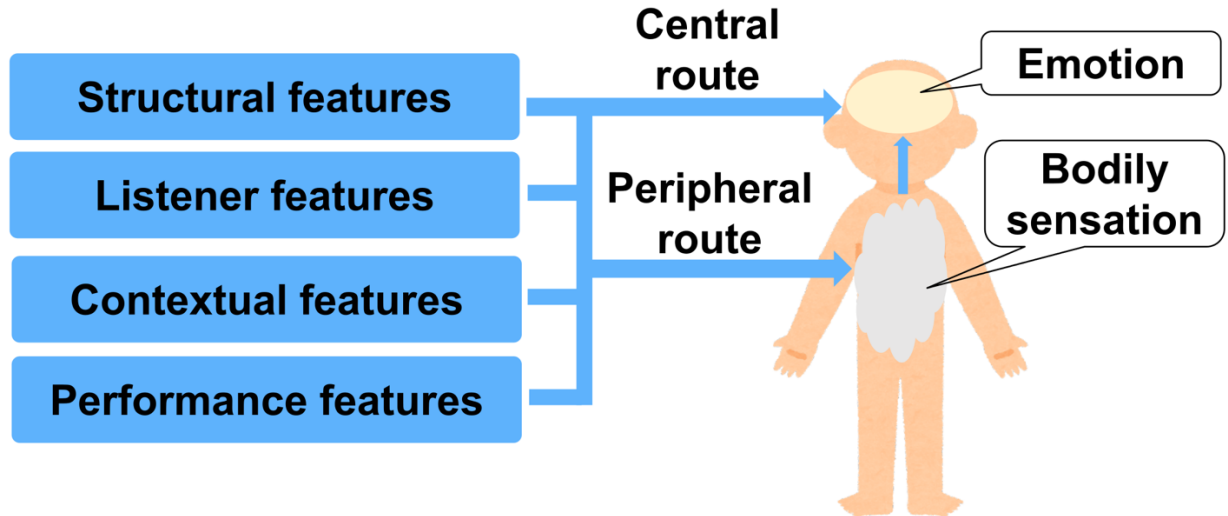

**Figure 1. Framework of emotion induction by auditory stimuli proposed by Scherer & Zentner (*2001*)**. Emotional experience is induced by four factors: structural features (acoustic and syntactic features of the auditory stimuli), listener features (listener's personal traits and transient states), contextual features (the listening environment and context), and performance features (the player's performance). These features evoke emotions directly through the brain (central route) and indirectly through activation of bodily sensations (peripheral route).

Previous studies on structural features have revealed the acoustic features that relate to felt valence (pleasantness) and arousal (emotion activation). Gómez et al. (*2007*) found that mode, harmonic complexity, and rhythmic articulation best distinguished pleasantness, while tempo, accentuation, and rhythmic articulation best distinguished arousal. Countinho et al. (*2011*) found that valence was positively associated with pitch and tempo, while arousal was positively associated with loudness, tempo, timbre, and pitch. Jaquet et al. (*2014*) found that lower pitch level was associated with more negative valence and higher arousal by using musical excerpts that were systematically varied in pitch level.

Previous studies on listener features have revealed that certain listener traits are strongly associated with auditory-induced emotion. Liljeström et al. (*2013*) found that listeners with higher openness to experience tended to experience more intense and positive emotions (such as happiness and pleasure). Gerstgrasser et al. (*2023*) revealed that musical expertise, listeners' personal traits (such as openness to experience), and current mood state are associated with enhanced intensity and differentiation (granularity) of the emotional experience. Sakka et al. (*2018*) suggested that severely depressed listeners tend to experience less happiness when listening to music that typically evokes happy memories. Kreutz et al (*2008*) found a correlation between absorption traits and emotional arousal. Rawlings et al. (*2008*) showed that psychoticism was associated with positive emotional responses to unpleasant music. Chang et al. (*2012*) pointed out that over- and under-responsiveness to auditory stimuli are related to problematic emotional and behavioral responses.

Studies have also investigated how bodily sensations in the peripheral route are associated with emotional responses to auditory stimuli. Dibben (*2004*) demonstrated that listening to music while in a physiologically aroused state enhances the intensity of the emotional experience. Coutinho et al. (*2011*) revealed that the listeners' emotion could be more accurately predicted by including physiological information obtained from skin conductance and heart rate.

Furthermore, studies have examined how bodily sensations influence emotional experience in conjunction with structural features of the sound input. Liljeström (*2013*) observed a high level of bodily arousal (measured by skin conductance and heart rate) and a high level of self-reported emotion intensity when participants listened to self-chosen music. Gomez et al. (*2007*) found that fast, accentuated, and staccato music induced faster breathing and higher minute ventilation, skin conductance, and heart rate. Putkinen et al. (*2024*) elucidated that acoustic features were linked with bodily sensations and emotion ratings across Western and East Asian cultures. Daikoku et al. (*2024*) provided insight into how musical expectation shapes emotional and physiological responses and revealed that chord progressions that shifted from low uncertainty and low surprise to low uncertainty and high surprise elicited cardiac sensations, which were associated with positive valence.

These findings have provided insight into the contributions of structural features, listener features, and bodily sensations to auditory-induced emotion. However, while structural features encompass both acoustic and syntactic features, prior research has primarily focused on acoustic features, leaving the study of syntactic features—organization of discrete structural elements into structured sequences (*Asano, 2015*)—largely unexplored.

Among such syntactic features are hierarchical syntactic structures—how elements of auditory stimuli are organized by local and nested global relationships (*Koelsch, 2013*). This structural complexity is a key component of human cognition in both music and language (*Koelsch, 2013*).

Previous research on hierarchical syntactic structures has been mainly focused on cognitive processing of such structures. Some of the main findings in the auditory modality include the preceding of global processing over local processing (*Sanders, 2007; List, 2007*). and the association between specific listener features, such as musical expertise and autism spectrum order, and enhanced local processing (*Ouiment, 2012; Susini, 2020; Mottron, 2000*).

In terms of the relationship between hierarchical syntactic structures and emotional experience, research is very limited. A few studies investigated the relationship in the visual modality and have suggested a bidirectional relationship between global processing and stronger happiness (*Ji, 2019; Gasper, 2002*). However, to the best of our knowledge, there are currently no studies that have investigated this relationship in the auditory modality. Moreover, there is no research on how bodily sensations affect emotions induced by hierarchical syntactic structures.

To address this gap, the current study investigated how the hierarchical syntactic structure of auditory stimuli influences emotional experience in conjunction with listener features and bodily sensations. To this end, we conducted an experiment where we created sound sequences that systematically varied in the complexity of the hierarchical structure and examined the subjective emotional and bodily response to those sequences. We hypothesized that variations in structural complexity will elicit distinct emotional responses in individuals with specific sensory processing patterns. Moreover, we predicted that bodily sensations would amplify those emotional responses, based on previous findings that bodily sensations enhanced emotion prediction (*Couthinho, 2011*).

## 2. Materials and Methods

**2.1. Participants**

715 Japanese adults (female = 357, male = 356, other = 2; mean age = 35.61 years, SD = 8.88) participated in the study. None of the participants had any hearing disorders or neurological disorders (based on self-report). All participants were recruited through an online survey company, Cross Marketing Inc. (*2024*), and were compensated for participation at a fixed rate determined by the company. The study was carried out in accordance with the Declaration of Helsinki and received approval from the Ethics Committee of The University of Tokyo (Approval No. UT-IST-RE-230601). Participants provided informed consent before starting the experiment, which they then completed online using a personal computer.

**2.2. Stimuli**

For auditory stimuli with varying structural complexity, we prepared sound sequences with pitch movement that followed varying global and local syntactic patterns.

*2.2.1. Sounds*

To form sound sequences, we first created eight sounds. Each sound was a Shepard tone (*Shepard, 1964*) with frequencies:

$$440 * 2^{\frac{i}{8}} \; Hz \; (i = 0, ..., 7)$$

All sounds had a duration of 400ms, a constant loudness, and a sampling frequency of 44,100 Hz.

*2.2.2. Sound Sequences*

Using the created sounds, we formed sound sequences with pitch movements following varying global and local syntactic patterns. We adopted a stimuli design used in previous work, which provides an auditory parallel to Navon's widely used visual local-global stimuli (*Navon, 1977*). The design is used in various global and local auditory processing studies (*Justus, 2005; List, 2007; Sanders, 2007; Ouimet, 2012; Bouvet; 2011*).

Each sound sequence consisted of two sets of four quadruplets. A quadruplet consisted of four of the created sounds which follow a local syntactic pattern. Four quadruplets were concatenated to follow a global syntactic pattern. The four-quadruplet sequence was repeated twice.

The local and global syntactic patterns were categorized into three types of pitch movements, each differing in complexity: static (unchanged), linear, and zigzag (Figure 2A). The static pattern was assigned a Complexity Level of 0, the linear pattern a Complexity Level of 1, and the zigzag pattern a Complexity Level of 2. Nine types of sound sequences were produced as a combination of the three levels of Global Complexity (Complexity Level of the global syntactic pattern between quadruplets: G0,

G1, G2) and the three levels of Local Complexity (Complexity Level of the local syntactic pattern within quadruplets: L0, L1, L2).

Furthermore, we created ascending and descending variations for the Complexity Level 1 and 2 patterns (linear and zigzag movement) (Figure 2A). Also, for the sound sequence with a combination of Global Complexity Level G0 and Local Complexity Level L0, we created two variations using different sounds in order to reduce a preference for a specific sound. As a result, a total of 26 sound sequences were created. A table of created sound sequence patterns is depicted in Figure 2B.

A 200-ms interval was placed between all sounds in the sequence to make the sounds seem more independent of each other, and hence the local and global syntactic patterns less obvious. The total duration of a sound sequence was 19 seconds.

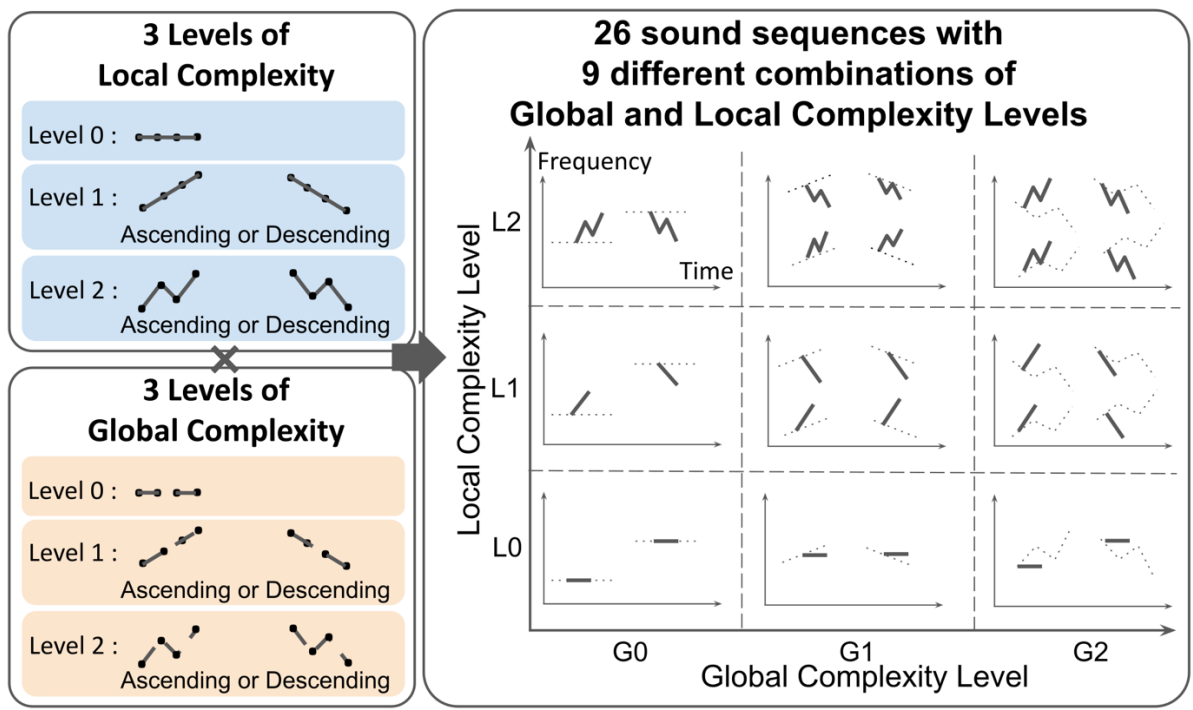

(A) Syntactic patterns for different complexity levels

(B) Sound sequences created by combining syntactic patterns of different complexity levels

**Figure 2. Sound sequences with systematically varied Local and Global Complexity.** **(A)** Sound sequences were created by combining three levels of Local and Global Complexity in pitch movement (unchanged, linear, zigzag) with two variations in pitch

direction (ascending, descending). **(B)** As a result, a total of 26 sound sequences across nine distinct complexity categories were produced.

### 2.3. Measures

#### *2.3.1. Experienced Emotion*

Emotions experienced by listening to the sound sequences were quantified with the two-dimensional model of valence and arousal proposed by Russel (*1980*). Valence and arousal are widely recognized as fundamental dimensions of emotions and have been shown to account for the majority of observed variance in the emotional labeling of several types of experimental stimuli, including linguistic, pictorial, and musical (*Gomez, 2007*; *Coutinho, 2011*). Valence and arousal were each rated on a 9-point Likert scale.

In addition to valence and arousal ratings, participants provided subjective ratings of noisiness and complexity of each sound sequence. Both were rated on a 4-point Likert scale. They also selected up to three emotional categories (from a list of 18, including options such as "happy" and "sad") that best described their emotional experience.

#### *2.3.2. Bodily Sensations*

Bodily sensations felt by listening to the sound sequences were measured by using an adaptation of the bodily map of emotion proposed by Nummenmaa et al. (*2014*). Participants were presented with a body image divided into 33 sections and were asked to select at least one section where they felt bodily sensation while they listened to the presented sound sequence (or where they would feel a bodily sensation if they continued to listen to) (Figure 3A).

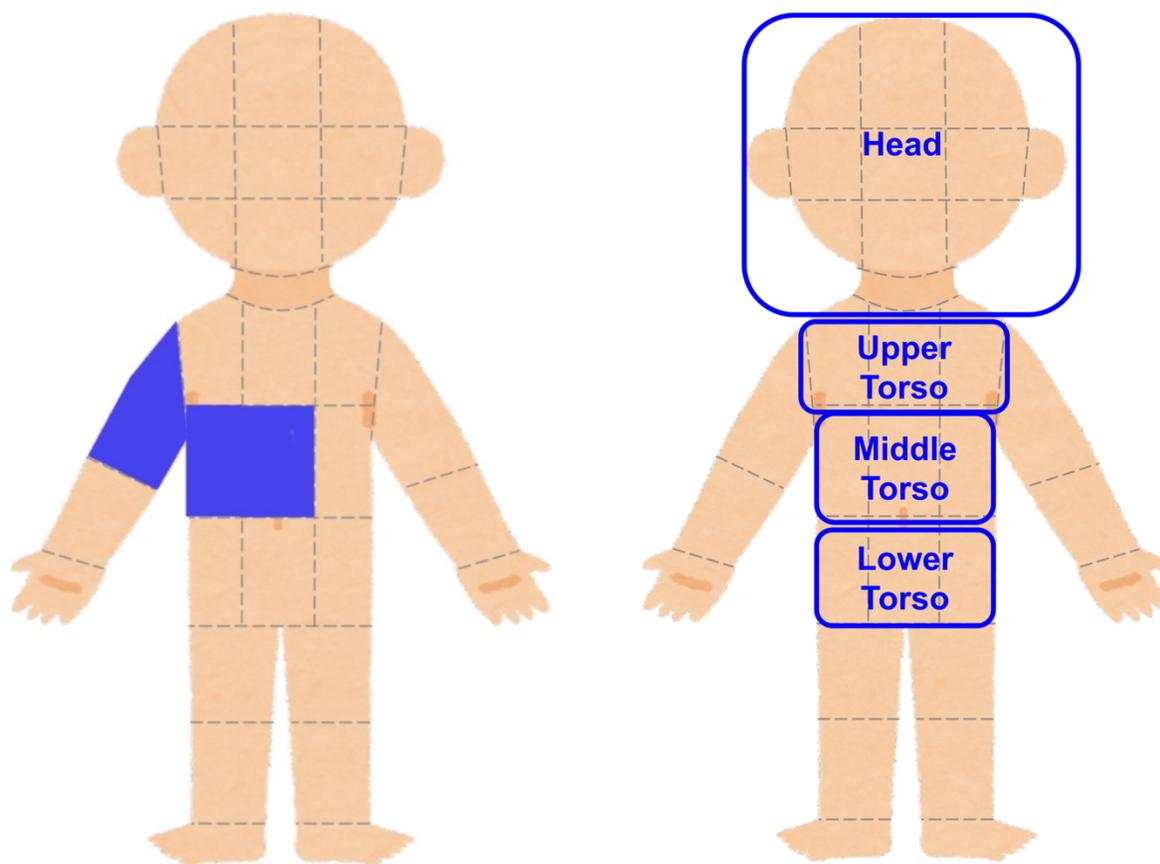

(A) Bodily map for reporting bodily sensations    (B) Grouping of body areas

**Figure 3. Bodily map and grouping of body sections. (A)** Participants clicked on the body sections where they felt (or expected to feel) bodily sensations while listening to the presented sound sequences. **(B)** The body sections were grouped into broader body areas which are suggested to be associated with emotional response to auditory stimuli.

### *2.3.3. Listener Features*

As listener features, we obtained the participants' sensory profile, autistic traits, intolerance to uncertainty, age, gender, and musical experience.

The sensory profile was assessed using the Japanese version of the Adolescent-Adult Sensory Profile (AASP) (*Brown, 2011*). The AASP is a 60-item questionnaire designed to evaluate sensory processing across six sensory domains: taste/smell, movement, visual, touch, activity level, and hearing. It provides scores for four sensory processing patterns, each representing a combination of neurological threshold and behavioral response to sensory input: Low Registration (high threshold x passive behavior), Sensation Seeking (high threshold x counteractive behavior), Sensory Sensitivity (low threshold x passive behavior), and Sensation Avoiding (low threshold x counteractive behavior). Each sensory processing pattern score ranges from 15 to 75, with higher scores indicating a stronger tendency toward the corresponding sensory processing pattern. We chose to use this measure because it captures both typical and atypical sensory processing patterns, accounting for listener characteristics such as age, but without focusing exclusively on any specific atypical condition or disorder to explain sensory processing patterns (*Dean, 2022*).

Autistic traits were assessed using the Japanese short form of the Autism Spectrum Quotient (AQ-J-10). The AQ-J-10 is a 10-item self-report questionnaire derived from Baron-Cohen et al.'s (*2006*) AQ questionnaire. It was developed by Kurita et al. (2005) for screening patients with high-functioning pervasive developmental disorder, now classified under autism spectrum disorder (*American Psychiatric Association, 2013*). The AQ-J-10 score ranges from 0 to 10, with higher scores indicating stronger autistic traits. Autistic traits are known to be linked to the processing of hierarchical structures (*Mottron, 2000; Sapey-Triomphe, 2023*).

Intolerance to uncertainty was assessed with the Japanese version of the Short Intolerance of Uncertainty Scale (SUIS), a 12-item questionnaire developed by Takebayashi et al. (*2012*) based on previous English versions (*Carleton, 2007; Freeston, 1994*). The SIUS score ranges from 12 to 60, with a higher score indicating greater intolerance to uncertainty. Uncertainty is known to be linked to processing of hierarchical structures and to ASD (*Hansen, 2014; Kluger 2020; Sapey-Triomphe, 2023, Vasa, 2018*).

Musical experience was quantified as the number of years spent receiving musical education or performing music outside the school curriculum. In addition, we obtained a brief description of the content of the musical experience.

**2.4. Procedures**

The experimental paradigm was created using the Gorilla Experiment Builder (*Anwyl-Irvine, 2020*), an online behavioral experiment builder. Participants accessed the experimental website using their personal computers from a location of their choice, were briefed on the details of the experiment in writing on the screen, consented to participation, completed screening questions, and started the experiment. The participants first completed the questionnaire on their listener features as mentioned above. Then they moved on to the stimuli rating task, where they were presented with the 8 single sounds and subsequently asked to provide their emotional response to each tone. Next, the 26 sound sequences were presented and participants were subsequently asked to provide their emotional and bodily response to each sound sequence. The single sounds and sound sequences were presented in randomized order for each participant.

## 2.5. Statistical Analyses

Statistical analyses were conducted using Jamovi Version 2.3 (*Şahin, 2019; The jamovi project, 2024*). Participants with missing responses were omitted prior to the analyses.

For each participant, the valence, arousal, noisiness, and complexity ratings were averaged across sound sequences with the same Local and Global Complexity Levels, producing nine averaged values for each rating (valence and arousal ratings for sound sequences with Complexity Level of L0G0, L0G1, L0G2, L1G0, L1G1, L1G2, L2G0, L2G1, L2G2).

We assessed the normality of valence and arousal ratings using the Shapiro-Wilk test, skewness and kurtosis values, as well as visual inspection of histograms and Q-Q plots. Based on these results, we decided whether to use either a parametric or non-parametric repeated-measures analysis of variance (ANOVA) to evaluate the effects of various factors on the valence and arousal ratings, respectively.

First, repeated-measures ANOVAs were conducted to examine the effects of structure features, listener features, and bodily sensations on valence ratings. In all ANOVAs, the dependent variable was the valence score, with two within-subject factors and two between-subject factors. The within-subject factors were: (1) the Local Complexity Level (L0, L1, L2) of the sound sequences, and (2) the Global Complexity Level (G0, G1, G2) of the sound sequences. The between-subject factors varied across ANOVAs and consisted of (1) the score of one AASP sensory processing pattern and (2) the presence or absence of bodily sensation in a specific body area. The specific AASP sensory processing patterns and bodily sensation areas used in the analysis are shown in Table 1. The thresholds for dividing participants into high and low groups for each sensory processing pattern were set to ensure an equal split, with half of the participants in each group. The body areas in the bodily sensation factors are depicted in Figure 3B. Participants who reported a bodily sensation in a specific body area for at least one sound sequence were classified into the group *with* bodily sensation in that body area. For example, a participant who reported a bodily sensation in the Head area for three sound sequences was categorized as a member of the *with bodily sensation in the Head* group. These four body areas were selected based on prior research suggesting connections between these areas and emotional responses (*Daikoku, 2024; Nummennmaa, 2014; Putkinen, 2024*).

Secondly, repeated-measures ANOVAs were conducted to evaluate the effects of structure features, listener features, and bodily sensations on arousal ratings. The dependent variables were arousal ratings. The within- and between-subject factors were the same as those used in the ANOVAs for valence ratings.

For all ANOVAs, the significance level was set at 5 %. We used post-hoc contrasts to follow up statistically significant interactions (and/or main effects with more than 2 levels), The p-values of post-hoc tests were adjusted based on the false discovery rate. Effect sizes were estimated using partial eta squared($\eta_p^2$).

**Table 1. Between-subject factors in ANOVA**

| Score of AASP sensory processing pattern | Presence of bodily sensation in body area |
|---|---|
| Low Registration (high vs. low) | Head (with vs. without) |
| Sensation Seeking (high vs. low) | Upper torso (with vs. without) |
| Sensory Sensitivity (high vs. low) | Middle torso (with vs. without) |
| Sensation Avoiding (high vs. low) | Lower torso (with vs. without) |

In addition, we conducted independent samples t-tests to examine the differences in AQ and SIUS scores in high and low score groups of each AASP sensory processing patterns. The significance level for the t-tests were set at 5%.

The collected emotion ratings for the single sounds and subjective ratings of noisiness, complexity, and emotion categories for the sound sequences were not included in the present analyses as they fall outside the scope of this study. Musical experience was also not included in the analyses due to the small number of participants with musical experience compared to those without. The complete raw dataset, including these additional data, is publicly available in the Raw_data folder at https://osf.io/tpqnj/files/osfstorage.

## 3. Results

A total of 579 participants (female = 283, male = 294, other = 2; mean age = 35.88 years, SD = 8.86) were included in the statistical analysis after excluding those with missing responses. The complete raw dataset for participants with complete responses, along with all analysis results, is publicly available at https://osf.io/tpqnj/.

### 3.1. ANOVA Results

We first examined the distribution of the valence and arousal ratings. The Shapiro-Wilk test showed significant deviations from normality ($p < 0.001$). However, given the large sample size (> 500) and a W statistic close to 1, these deviations were not considered problematic. Additionally, skewness and kurtosis fell within the acceptable range for normality (between -2 and +2) (*George, 2010*). Considering these factors along with visual inspection of histograms and Q-Q plots, we assumed that the data could be sufficiently approximated by a normal distribution (histograms and Q-Q plots can be found in the Analysis_results folder at https://osf.io/tpqnj/files/osfstorage). Based on this assumption, we conducted the parametric repeated-measures ANOVA on the dataset.

#### *3.1.1. Valence*

The main effect of Local Complexity Level was significant across all combinations of sensory processing pattern scores and bodily sensation areas used as between-subject factors. Taking the Sensory Sensitivity score and Middle Torso area combination as an example, the main effect of Local Complexity Level was $F(2,1150) = 6.54$ ($p = 0.001$, $\eta_p^2 = 0.011$). Post-hoc tests revealed that valence ratings of sound sequences with Local Complexity Level of L1 (medium) were significantly higher than those of L0 (low) and L2 (high) (L0: $p = 0.095$, L2: $p = 0.003$) (Supplementary Figure 1). The L1 valence was significantly higher than L2 valence for all sensory processing patterns and bodily sensation area combinations ($p < 0.04$).

The main effect of Global Complexity Level was also significant across all combinations of sensory processing pattern scores and body sensation area. Again, for the Sensory Sensitivity score and Upper Torso area combination, the main effect of Global Complexity Level was significant with $F(2,1150) = 72.40$ ($p < 0.001$, $\eta_p^2 = 0.112$). Post-hoc tests revealed that the valence ratings of sound sequences with Global Complexity Level of G0 (low) were lower than those of both G1 and G2 (medium), regardless of participants' sensory processing patterns or bodily sensations ($p < 0.002$) (Supplementary Figure 2).

The within-subject interactions of Local Complexity Level and Global Complexity Level were also significant across all sensory processing pattern scores x body area combinations. As an example, for the Sensory Sensitivity score and Upper Torso area combination, the interaction showed $F(4,2300) = 57.01$ ($p < 0.001$, $\eta_p^2 = 0.090$) (Figure 4). Post-hoc tests revealed that, among sound sequences with Global Complexity Level of G0 (low), the valence ratings were lowest for sequences with Local Complexity Level of L0 (low). Among sound sequences with Global Complexity Level of G1 (medium), the valence ratings were the lowest for sequences with Local Complexity Level of L2 (high). Among sound sequences with Global Complexity Level of G2 (high), the valence ratings were the highest for sequences with Local Complexity Level of L0 (low). Additionally, the L0xG0 sound sequences (lowest local and global complexity) received the lowest overall valence ratings. These observations were significant across all sensory processing pattern score and bodily sensation area combinations.

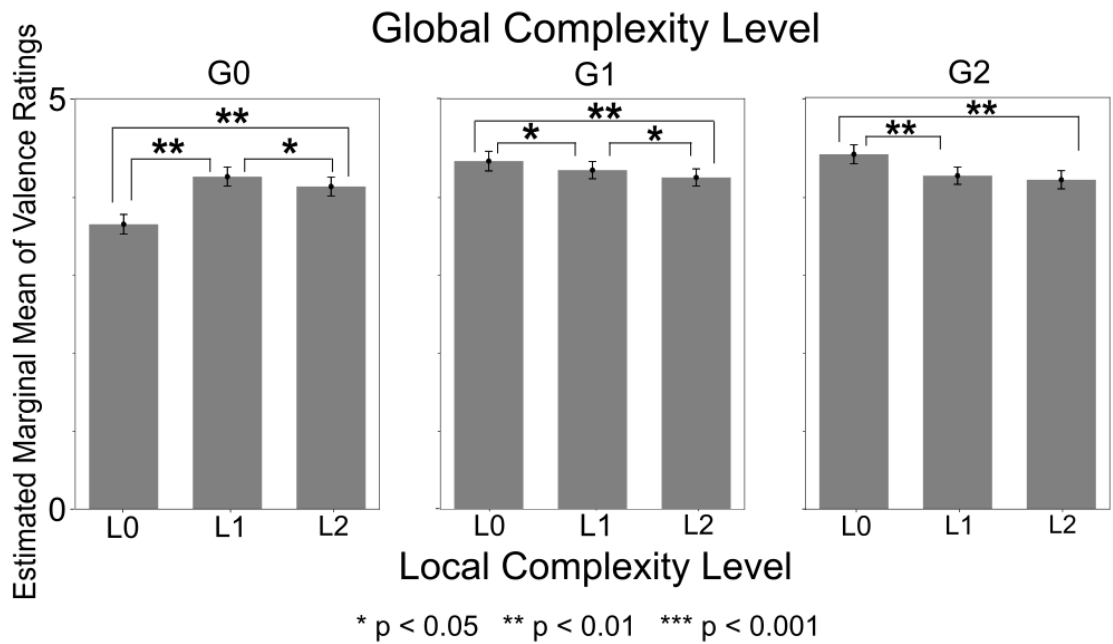

**Figure 4. Valence Ratings by Local Complexity Level x Global Complexity Level.** Valence ratings were significantly higher for sound sequences with high Local and low Global Complexity (L1xG0, L2xG0) compared to sequences with low Local and low Global Complexity (L0xG0). In contrast, valence ratings were higher for sequences with low Local and medium or high Global Complexity (L0xG1, L0xG2) compared to sequences with high Local and medium or high Global Complexity (L1xG1, L2xG1, L1xG2, L2xG2).

The ANOVA with Sensory Sensitivity score and Upper Torso area (with vs. without Upper Torso sensation) as between-subject factors revealed significant main effects for both Sensory Sensitivity score ($F(1,575) = 4.56$, $p = 0.033$, $\eta_p^2 = 0.008$) and Upper Torso area ($F(1,575) = 11.27$, $p < 0.001$, $\eta_p^2 = 0.019$). However, the interaction between Sensory Sensitivity score and Upper Torso area was insignificant ($p = 0.852$). Local Complexity Level x Global Complexity Level x Sensory Sensitivity score x Upper Torso area interaction was also significant ($F(4,2300) = 2.48$, $p = 0.042$, $\eta_p^2 = 0.004$). Post-hoc comparisons revealed that valence ratings were higher in the low Sensory Sensitivity score group than in the high Sensory Sensitivity score group, and higher in participants who reported Upper Torso bodily sensations compared to those who did not.

Within the high Sensory Sensitivity score group, participants who experienced Upper Torso sensations showed significantly higher valence ratings for sequences with higher combinations of Local and Global Complexity Levels (L1xG1, $p = 0.014$; L1xG2, $p = 0.028$; L2xG2, $p = 0.010$) compared to those who did not experience Upper Torso sensation (Figure 5A). Furthermore, the decline in valence ratings for G2 sequences as Local Complexity Level increased was significantly smaller in participants with Upper Torso sensations than in those without. Specifically, among participants *without* Upper Torso sensation, the mean valence difference between L0xG2 and L1xG2 sequences was 0.28 ($p = 0.011$), and between L0xG2 and L2xG2 sequences was 0.39 ($p = 0.011$). In contrast, among participants *with* Upper Torso sensations, no significant differences in valence were observed between L0xG2 and L1xG2 sequences ($p = 0.208$) or L0xG2 and L2xG2 sequences ($p = 0.142$).

Conversely, within the low Sensory Sensitivity score group, participants who experienced Upper Torso sensations showed significantly higher valence ratings for sequences with lower Local and higher Global Complexity Levels (L0xG1, $p = 0.031$; L0xG2, $p = 0.015$) compared to those who did not experience Upper Torso sensation (Figure 5B). Furthermore, the decline in valence ratings of G1 and G2 sequences as Local Complexity Level increased was significantly larger in participants with Upper Torso sensation compared to those without. Specifically, in participants *with* Upper Torso sensations, the mean valence difference between L0xG1 and L2xG1 sequences was 0.34 ($p = 0.015$), and between L0xG2 and L2xG2 sequences was 0.47 ($p = 0.011$). In contrast, among participants *without* Upper Torso sensations, the mean valence difference between L0xG1 and L2xG1 sequences was 0.19 ($p = 0.037$), and between L0xG2 and L2xG2 sequences was 0.21 ($p = 0.015$).

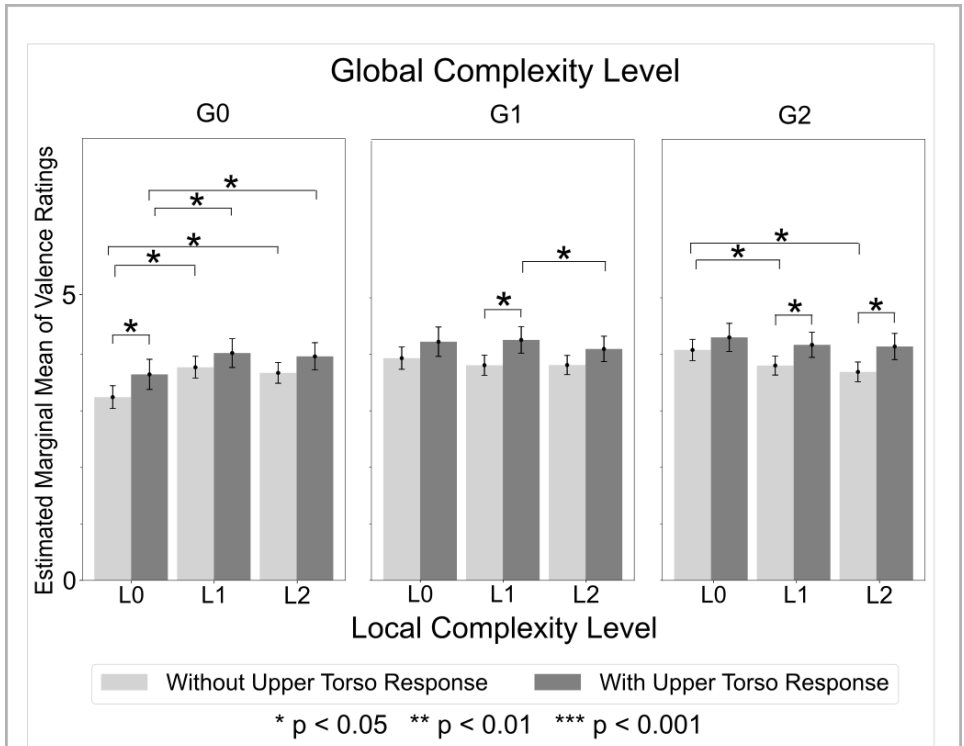

(A) High Sensory Sensitivity score group

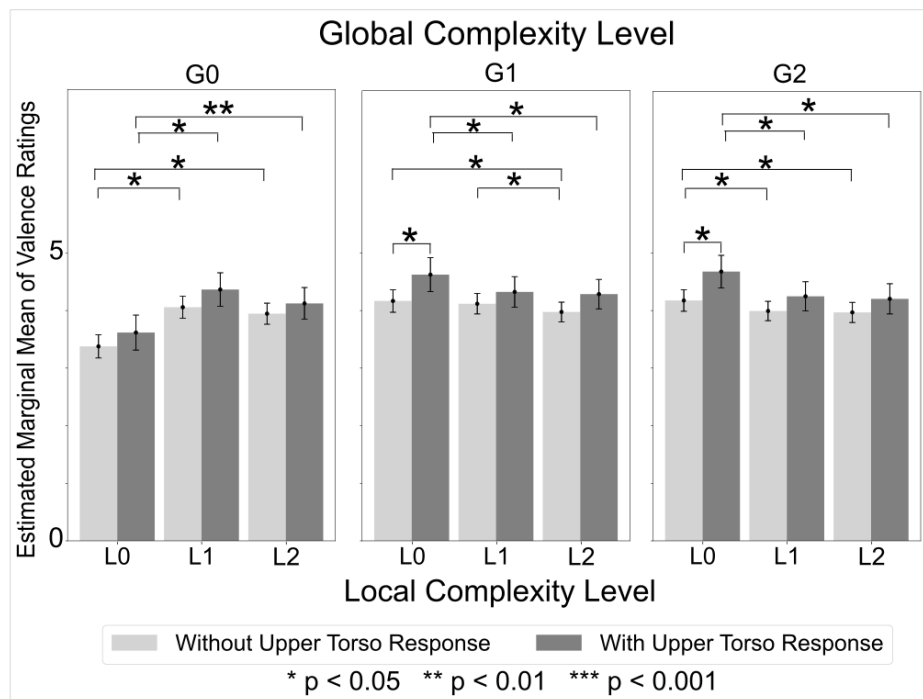

(B) Low Sensory Sensitivity score group

**Figure 5. Valence Ratings by Local Complexity Level x Global Complexity Level in high and low Sensory Sensitivity score groups. (A)** In the high Sensory Sensitivity score group, participants who experienced Upper Torso sensations showed significantly higher valence ratings for sequences with higher combinations of Local and Global Complexity levels compared to those who did not experience Upper Torso sensations. **(B)** By contrast, in the low Sensory Sensitivity score group, participants who experienced Upper Torso sensations showed significantly higher valence ratings for sequences with lower Local and higher Global Complexity levels.

Similarly, ANOVA with Sensory Sensitivity score and Middle Torso area as between-subject factors revealed significant main effects for both Sensory Sensitivity score ($F(1,575) = 5.73$, $p = 0.017$, $\eta_p^2 = 0.01$) and Middle Torso area ($F(1,575) = 7.11$, $p = 0.008$, $\eta_p^2 = 0.012$), and a non-significant Sensory Sensitivity x Middle Torso interaction ($p = 0.652$). The Local Complexity Level x Global Complexity Level x Sensory Sensitivity score x Middle Torso area interaction was also significant ($F(4,2300) = 2.42$, $p = 0.047$, $\eta_p^2 = 0.004$). Post-hoc comparisons revealed that valence ratings were higher in the low Sensory Sensitivity score group compared to the high Sensory Sensitivity score group. Post-hoc comparisons also revealed that valence ratings were higher in participants with Middle Torso bodily sensation compared to participants without. Among participants with high Sensory Sensitivity scores, those who experienced sensations in the Middle Torso showed higher valence scores than those who did not, although the difference was not statistically significant.

Other significant effects and interactions observed were as follows: ANOVA with Sensation Avoiding score and Head area as between-subject factors revealed significant main effects for Sensation Avoiding scores ($F(1,575) = 5.64$, $p = 0.018$, $\eta_p^2 = 0.01$) and Head area ($F(1,575) = 4.19$, $p = 0.041$, $\eta_p^2 = 0.007$), with low Sensation Avoiding scores and presence of Head sensations yielding higher valence scores. ANOVA with Sensation Avoiding score and Middle Torso area as between-subject factors revealed significant main effects for Sensation Avoiding score ($F(1,575) = 6.41$, $p = 0.012$, $\eta_p^2 = 0.011$) and Middle Torso area ($F(1,575) = 7.01$, $p = 0.008$, $\eta_p^2 = 0.012$), with low Sensation Avoiding scores and presence of Middle Torso sensations yielding higher valence scores. ANOVA with Sensation Seeking scores and Upper Torso area as between-subject factors revealed significant main effects for Sensation Seeking score ($F(1,575) = 6.83$, $p = 0.009$, $\eta_p^2 = 0.012$) and Upper Torso area ($F(1,575) = 8.26$, $p = 0.004$, $\eta_p^2 = 0.014$), with high Sensation Seeking scores and presence of Upper Torso sensations yielding higher valence scores. The sensory processing pattern x bodily sensation area interactions were insignificant in all above combinations.

### *3.1.2. Arousal*

The main effect of Local Complexity Level was significant across all combinations of sensory processing pattern scores and bodily sensation areas used as between-subject factors. F-statistic was descriptively largest in the Sensation Avoiding scores and Upper Torso area combination ($F(2,1150) = 16.33$, $p < 0.001$, $\eta_p^2 = 0.028$). Post-hoc tests revealed that arousal was highest for L0 sound sequences (lowest Local Complexity Level), regardless of participants' sensory processing patterns or bodily sensations ($p \leq 0.003$) (Supplementary Figure 3).

The main effect of Global Complexity Level was also significant across all combinations of sensory processing pattern scores and body sensation area. F statistic was descriptively largest in the Sensation Seeking score and Upper Torso area combination ($F(2,1150) = 31.55$, $p < 0.001$, $\eta_p^2 = 0.052$). Post-hoc tests revealed that arousal was highest for G0 sound sequences (lowest Global Complexity Level), regardless of participants' sensory processing patterns or bodily sensations ($p < 0.002$) (Supplementary Figure 4).

The within-subject interaction of Local Complexity Level and Global Complexity Level was also significant across all combinations of sensory processing pattern scores and bodily sensation areas. The descriptively largest F-statistic was observed for the Sensation Seeking score and Upper Torso area combination ($F(2,1150) = 22.59$, $p < 0.001$, $\eta_p^2 = 0.038$). For the Sensation Avoiding score and Upper Torso area combination, the interaction was also significant ($F(4,2300) = 22.36$, $p < 0.001$, $\eta_p^2 = 0.037$). Post-hoc tests revealed that, regardless of participants' sensory processing patterns or bodily sensations, L0xG0 sound sequences (lowest Local and Global Complexity Levels) elicited significantly higher arousal compared to sequences with other Lower x Global Complexity combinations, as shown in Figure 6 ($p = 0.002$).

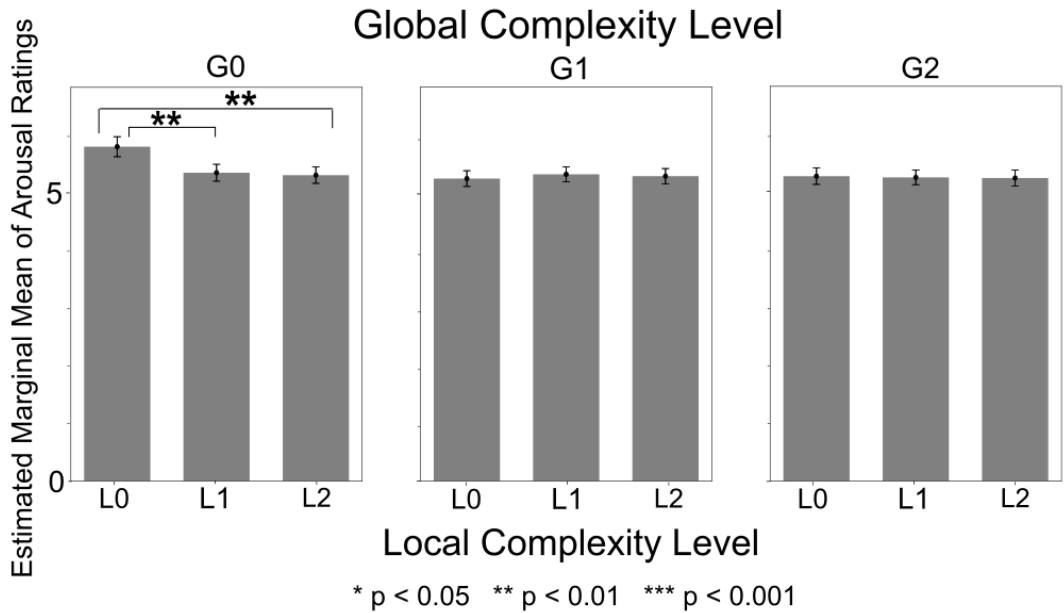

**Figure 6. Arousal Ratings by Local Complexity Level x Global Complexity Level.** The sound sequences with lowest Local and lowest Global Complexity (L0xG0) elicited significantly higher arousal compared to sequences with other Lower x Global Complexity combinations.

ANOVAs with Sensation Avoiding scores and body areas as between-subject factors revealed a significant main effect of Sensation Avoiding scores across all body area combinations, with the descriptively largest F-statistic found for the Sensation Avoiding score and Middle Torso area combination ($F(1,575) = 6.12$, $p = 0.014$, $\eta_p^2 = 0.011$). Post-hoc comparisons showed that arousal ratings were higher in the high Sensation Avoiding score group compared to the low Sensation Avoiding score group.

ANOVA with Sensation Seeking score and Upper Torso area as between-subject factors revealed a significant Global Complexity Level x Sensory Sensitivity score x Upper Torso area interaction ($F(2,1150) = 7.91$, $p < 0.001$, $\eta_p^2 = 0.014$). Post-hoc tests revealed that, within the low Sensation Seeking group, participants who experienced Upper Torso sensation exhibited a significantly greater decrease in arousal ratings as Global Complexity Level increased, compared to participants who did not experience Upper Torso sensations (Figure 7). Specifically, among participants with Upper Torso sensations, mean arousal ratings for G1 and G2 sequences were 0.36 and 0.46 lower than that for G0 sequences ($p = 0.009$). In contrast, among participants without Upper Torso sensations, the corresponding differences were only 0.15 and 0.15 ($p = 0.010$). Consequently, the arousal ratings for G1 and G2 sequences were significantly lower in participants with Upper Torso sensations compared to those without (G1: *p* = 0.041, G2: *p* = 0.01). However, no such significant mean differences were found within the high Sensation Seeking group.

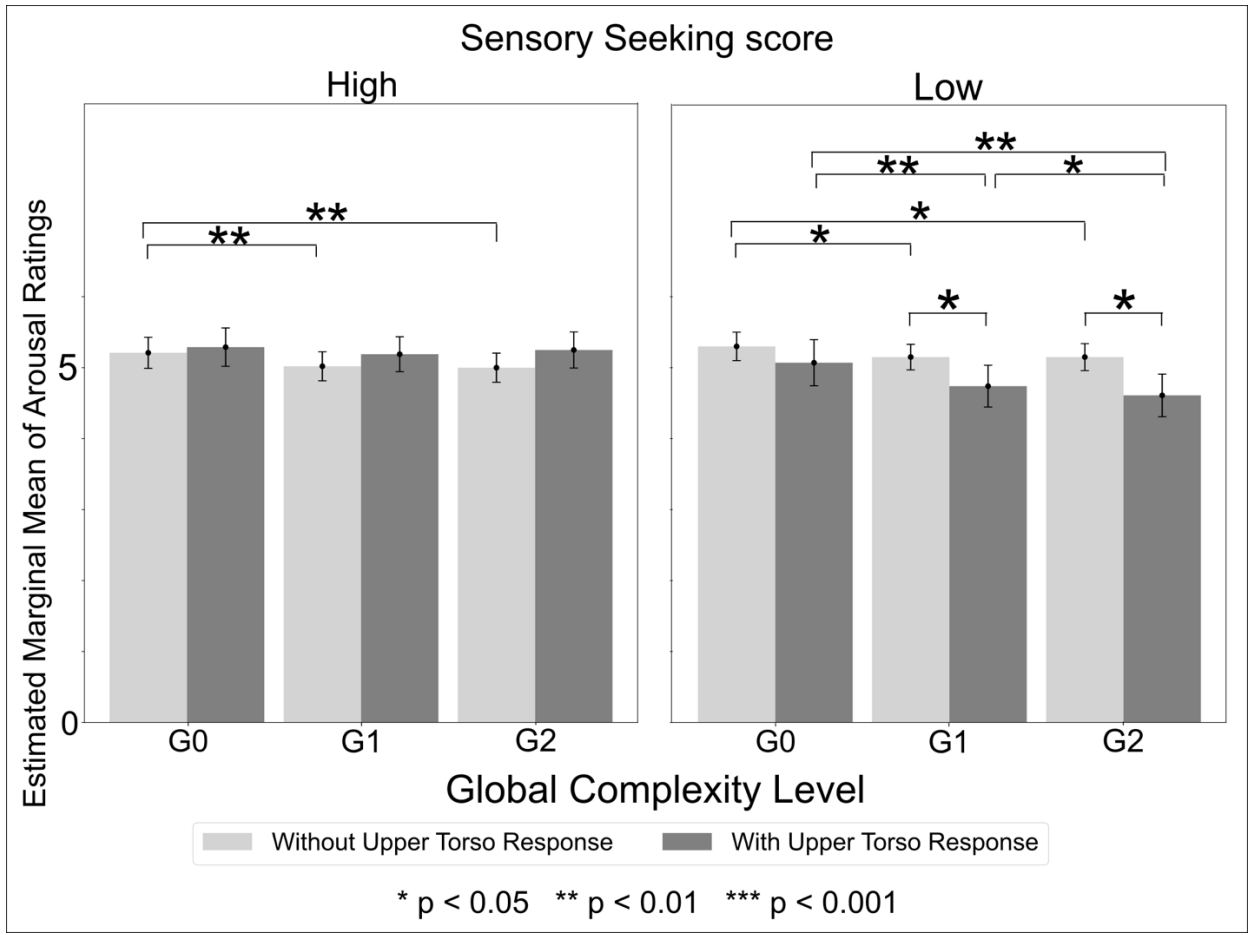

**Figure 7. Arousal Ratings by Global Complexity Level in high and low Sensation Seeking score groups.** In the low Sensation Seeking group, participants who experienced Upper Torso sensation showed a significantly greater decrease in arousal ratings as Global Complexity Level increased, compared to those who did not experience Upper Torso sensations. However, no significant differences were found within the high Sensation Seeking group.

For interactions of Global Complexity Level x Sensation Seeking score x Middle Torso area, Global Complexity Level x Sensory Sensitivity score x Middle Torso area, and Global Complexity Level x Low Registration score x Middle Torso, participants with low sensory processing pattern scores who experienced bodily sensations consistently showed a greater decrease in arousal scores as Global Complexity Level increased from G0 to either G1 or G2.

### 3.2. T-test Results

Welch's independent t-tests were conducted due to significant results of Levene's test for equality of variances for some AQ and SIUS scores. Participants with high Sensory Sensitivity scores had significantly higher AQ scores ($M$ = 4.63, $SD$ = 2.31) than those with low Sensory Sensitivity scores ($M$ = 3.15, $SD$ =1.84), $t(566)$ = 8.52, $p < 0.001$, Cohen's $d$ = 0.71. Similarly, participants with high Sensory Sensitivity scores had significantly higher SIUS scores ($M$ = 36.99, $SD$ = 7.45) than those with low Sensory Sensitivity scores ($M$ = 31.47, $SD$ = 7.35), $t(574)$ = 8.97, $p < 0.001$, Cohen's $d$ = 0.75.

A similar pattern was observed when participants were grouped by Sensation Avoiding and Low Registration scores, with both showing significant differences ($p < 0.001$). However, no significant differences were found between AQ or SIUS scores for participants with high versus low Sensation Seeking scores (AQ: $p$ = 0.45, SIUS: $p$ = 0.93) These results are depicted in Supplementary Figures 5-8.

## 4. Discussion

The current study aimed to elucidate how emotional experiences are influenced by the global and local syntactic structure of auditory sequences in corroboration with listeners' sensory processing patterns and auditory-induced bodily sensations. Participants were presented with sound sequences which had various levels of local and global complexity in their syntactic patterns. Emotions induced by the sound sequences were quantified by valence and arousal ratings.

Experimental results showed that emotional experience induced by a sound sequence is influenced by its hierarchical structure. Valence was higher for high Local x low Global Complexity and low Local x high Global Complexity structures. This result indicates that listeners preferred sound sequences with hierarchical structures that are in the Goldilocks zone, in other words, "not too simple, not too complex".

Meanwhile, arousal was highest for lowest Local x lowest Global Complexity structure, which received the lowest valence rating across all structures. Sound sequences of this structure were a repetition of same sound, sounding like the beeping of an alarm. This result suggests that such sequences were highly alarming but unpleasant to the listeners.

Results showed that listener's sensory processing patterns influence emotional experience. Participants with higher Sensory Sensitivity and Sensation Avoiding scores reported lower valence across all sound sequences compared to those with lower scores. Moreover, participants with higher Sensation Avoiding scores reported higher arousal. These results suggest that individuals with low neurological thresholds for stimuli are more likely to experience negative feelings when exposed to auditory stimuli and such feelings are more strongly activated in individuals who exhibit a tendency to avoid sensory stimuli. On the other hand, participants with higher Sensation Seeking scores reported higher valence compared to those with lower scores. This suggests that participants who crave sensory stimulation tend to feel more pleasure when they are exposed to auditory stimuli.

Further results showed that bodily sensation, especially in the Upper Torso area, also has influence on emotional experience. In terms of valence, within participants with high Sensory Sensitivity, those who experienced Upper Torso sensation felt (1) higher valence in sound sequences with *higher* Local and higher Global Complexity and (2) significantly *decreased* valence difference between low Local x high Global and high Local x high Global Complexity sequences, whereas those who didn't experience Upper Torso sensation felt (1) higher valence in sound sequences with *lower* Local and higher Global Complexity and (2) significantly *increased* valence difference between low Local x high Global and high Local x high Global Complexity sequences. In terms of arousal, within the low Sensation Seeking group, participants who experienced Upper Torso sensation exhibited a significantly greater decrease in arousal as the sound sequences' Global Complexity Level increased. These results indicate that an interplay of the sound sequence's hierarchical structure (Local and Global), listener's sensory processing pattern (Sensory Sensitivity and Sensation Seeking), and presence of Upper Torso sensation leads to different outcomes in emotional experience (valence or arousal). The results also suggest that the Upper Torso area may be a key body area in auditory-induced emotional experience.

The results also suggest that the AASP sensory processing patterns overlap with autistic traits and intolerance to uncertainty, both of which have been linked to hierarchical structures, as mentioned earlier. Furthermore, the results indicate that the AASP sensory processing patterns capture Sensation Seeking traits, which were not reflected in the AQ scores but are known to be frequently observed in individuals with ASD (*MacLennan, 2022*). Therefore, the AASP sensory processing patterns can be considered to adequately capture listener features that are closely related to hierarchical syntactic structures.

The current study supports the previously proposed framework which suggests that emotional experiences are triggered by a combination of factors, including audio structural features and listener characteristics through the central and peripheral routes (*Scherer & Zentner, 2001*). The current study extends the framework by elucidating how hierarchical syntactic structure (global and local syntactic patterns) influence emotional experience. Specifically, we showed that a "not too simple, not too complex" structure is most preferred, as indicated by higher valence ratings. This is in accordance with the widely accepted inverted-U shape relationship between pleasure and stimulus complexity (*Berlyne, 1971*). The inverted-U shape relationship in audition was found in studies such as the one by Cheung et al. (*2019*), whose findings indicated that chords with low uncertainty and high surprise or with high uncertainty and low surprise were most pleasurable. Reber et al. (*2004*) explained the inverted-U preference through processing fluency, proposing that ease of processing leads to positive affect. They suggested that structures initially perceived as complex but ultimately easy to process evoke particularly strong pleasure. Applying this concept, our findings can be interpreted as follows: In high Local Complexity x low Global Complexity structures, high Local Complexity creates an impression of difficulty, while low Global Complexity facilitates actual processing. Conversely, in low Local Complexity x high Global Complexity structures, low Local Complexity makes the structure simpler, but high Global Complexity introduces processing difficulty.

The current study also adds to the previous auditory-emotion framework by clarifying listener features that have stronger relationships with auditory-induced emotions. Specifically, our study suggests that high Sensory Sensitivity reduces positive emotions towards auditory stimuli, which is in line studies which indicate that individuals with high sensory sensitivity tend to be overwhelmed by environmental stimuli (*Harrold, 2024*).

Further, our study supports previously proposed connections between bodily sensation and auditory-induced emotion (*James, 1884; Scherer & Zentner, 2001*). Our study confirms the importance of the Upper Torso area, which covers the heart and stomach regions. Moreover, and most significantly, our study suggests that the link between bodily sensation and emotional experience depends on a complex interplay with the audio's hierarchical structure and listener's sensory processing pattern. Specifically, our findings showed that the presence of Upper Torso sensation alters valence, but the effect differs by the Local and Global complexity of the audio structure and how strong the listener's Sensory Sensitivity or Sensation Seeking trait is. Previous studies have demonstrated the influence of bodily sensations on emotion, particularly sensation in body parts included in the Upper Torso area, but their findings vary depending on other factors of emotion induction. For instance, Daikoku et al. (*2024*) found that heart-related sensations induced by an interplay of musical uncertainty and prediction error in musical chords were positively correlated with positive valence, whereas Putkinen et al. (*2024*) demonstrated the close link between chest areas and negative emotions with the auditory stimuli's acoustic features coming into play. Taken together, these findings highlight that bodily sensations alone are not deterministic; instead, their emotional impact is shaped by a complex interplay with the auditory stimuli's structure feature and listener feature. This underscores the need for an integrative approach in understanding the role of bodily sensations in emotion.

Our findings of the complex interplay of auditory hierarchical structure, listener sensory processing patterns, and bodily sensations have some possible clinical implications. In particular, the findings can be applied to enhance emotion regulation strategies in response to auditory stimuli. Firstly, they can help raise awareness of emotional reactions triggered by auditory stimuli with different hierarchical structures, allowing to avoid structures that are more likely to evoke negative emotions. Secondly, these findings open the possibility of developing new therapeutic approaches to regulate auditory-induced emotions by leveraging the control of bodily sensations. For example, developing a method to intentionally induce bodily sensation in the Upper Torso body area may help individuals with stronger Sensory Sensitivity traits feel fewer negative feelings induced by high Local and Global Complexity stimuli or help individuals with stronger Sensation Seeking traits feel less activation induced by higher Global Complexity stimuli, leading to a calmer mental state in both cases.

Although our study suggests a connection between the Upper Torso body area and auditory-induced emotions, the causal relationship remains unclear. Also, it is not clear whether the bodily sensations reported by participants were only subjective feelings ("the participants felt like they were feeling some bodily sensation") or involved actual physiological changes in the respective body area. This could be addressed in future experiments where we record both subjective emotion ratings and physiological bodily changes induced by auditory sequences.

While we did observe various significant effects, the effect sizes of the factors tested in the ANOVA of our current study were relatively small. This may be due to the influence of other well-established acoustic features, such as pitch and loudness, which are already known to significantly affect emotions but were not included in the current study. To accurately judge the impact of our current factors, one would need to analyze and compare the effect sizes of our current factors and such well-established factors together. Conducting a study that systematically incorporates factors identified as relevant in previous separate studies should be included in future works to provide a more comprehensive analysis of each factor's individual impact.

Our study was one of the first to explore the association between hierarchical structure and emotion in audition. While we were able to show how the global and local syntactic patterns are related to emotional experience, higher-level structures created by incorporating patterns, such as the nesting of ascending/descending pitch movement at certain locations of the sound sequence, are yet to be explored. Such higher-level hierarchical structures are suggested to evoke tension and relaxation (*Jackendoff, 2006*). Investigating emotional changes due to such higher-order hierarchical structures are expected to reveal further important findings.

In conclusion, the present study demonstrated that a complex interplay of audio structure feature, listener feature, and bodily sensations influence emotional experience induced by auditory stimuli. Findings provide empirical support for theories claiming multifactor triggers of auditory-induced emotion and connection between bodily sensation and emotion. They also suggest possible application in emotion regulation strategies. Potential directions of future research include measuring physiological signals to investigate the causal relationship between bodily sensation and emotion and integrating all previously identified modulators into one study to comprehensively investigate the relevant emotion triggers.

*Supplementary Materials for :*

**Emotional Responses to Auditory Hierarchical Structures is Shaped by Bodily Sensations and Listeners' Sensory Traits**

Maiko Minatoya, Tatsuya Daikoku, and Yasuo Kuniyoshi

Graduate School of Information Science and Technology, The University of Tokyo, Tokyo, Japan

1. Supplementary Tables

### Supplementary Table 1. Descriptives of Valence Ratings

#### (A) Shapiro-Wilk test results of Valence Ratings

| Local x Global complexity | Shapiro-Wilk W | Shapiro-Wilk p |
|---|---|---|
| L0xG0 | 0.964 | <.001 |
| L0xG1 | 0.965 | <.001 |
| L0xG2 | 0.969 | <.001 |
| L1xG0 | 0.973 | <.001 |
| L1xG1 | 0.977 | <.001 |
| L1xG2 | 0.973 | <.001 |
| L2xG0 | 0.975 | <.001 |
| L2xG1 | 0.979 | <.001 |
| L2xG2 | 0.979 | <.001 |

#### (B) Skewness and kurtosis of Valence Ratings

| Local x Global complexity | Skewness | Std. error skewness | Kurtosis | Std. error kurtosis |
|---|---|---|---|---|
| L0xG0 | 0.00824 | 0.102 | -0.586 | 0.203 |
| L0xG1 | -0.373 | 0.102 | -0.196 | 0.203 |
| L0xG2 | -0.312 | 0.102 | 0.272 | 0.203 |
| L1xG0 | -0.21 | 0.102 | -0.0431 | 0.203 |
| L1xG1 | -0.362 | 0.102 | 0.0757 | 0.203 |
| L1xG2 | -0.475 | 0.102 | -0.0205 | 0.203 |
| L2xG0 | -0.13 | 0.102 | -0.0615 | 0.203 |
| L2xG1 | -0.379 | 0.102 | 0.0279 | 0.203 |
| L2xG2 | -0.345 | 0.102 | -0.0643 | 0.203 |

**(C) Mean and Median of Valence Ratings**

| Local x Global complexity | Mean | Median | Standard deviation |
|---|---|---|---|
| L0xG0 | 3.42 | 3.5 | 1.43 |
| L0xG1 | 4.17 | 4.5 | 1.4 |
| L0xG2 | 4.24 | 4.5 | 1.34 |
| L1xG0 | 4 | 4 | 1.38 |
| L1xG1 | 4.07 | 4.25 | 1.26 |
| L1xG2 | 4 | 4 | 1.19 |
| L2xG0 | 3.88 | 4 | 1.3 |
| L2xG1 | 3.99 | 4 | 1.21 |
| L2xG2 | 3.94 | 4 | 1.25 |

## Supplementary Table 2. Descriptives of Arousal Ratings.

### (A) Shapiro-Wilk test results of Arousal Ratings

| Local x Global complexity | Shapiro-Wilk W | Shapiro-Wilk p |
|---|---|---|
| L0xG0 | 0.974 | <.001 |
| L0xG1 | 0.971 | <.001 |
| L0xG2 | 0.965 | <.001 |
| L1xG0 | 0.972 | <.001 |
| L1xG1 | 0.973 | <.001 |
| L1xG2 | 0.974 | <.001 |
| L2xG0 | 0.975 | <.001 |
| L2xG1 | 0.979 | <.001 |
| L2xG2 | 0.975 | <.001 |

### (B) Skewness and Kurtosis of Arousal Ratings

| Local x Global complexity | Skewness | Std. error skewness | Kurtosis | Std. error kurtosis |
|---|---|---|---|---|
| L0xG0 | -0.185 | 0.102 | -0.187 | 0.203 |
| L0xG1 | 0.162 | 0.102 | 0.533 | 0.203 |
| L0xG2 | 0.123 | 0.102 | 0.65 | 0.203 |
| L1xG0 | 0.0612 | 0.102 | 0.369 | 0.203 |
| L1xG1 | 0.151 | 0.102 | 0.932 | 0.203 |
| L1xG2 | 0.19 | 0.102 | 0.71 | 0.203 |
| L2xG0 | 0.068 | 0.102 | 0.326 | 0.203 |
| L2xG1 | 0.172 | 0.102 | 0.678 | 0.203 |
| L2xG2 | 0.154 | 0.102 | 0.676 | 0.203 |

**(C) Mean and Median of Arousal Ratings**

| Local x Global complexity | Mean | Median | Standard deviation |
|---|---|---|---|
| L0xG0 | 5.54 | 5.5 | 1.88 |
| L0xG1 | 5.03 | 5 | 1.51 |
| L0xG2 | 5.07 | 5 | 1.51 |
| L1xG0 | 5.11 | 5 | 1.59 |
| L1xG1 | 5.09 | 5 | 1.39 |
| L1xG2 | 5.04 | 5 | 1.42 |
| L2xG0 | 5.07 | 5 | 1.56 |
| L2xG1 | 5.07 | 5 | 1.43 |
| L2xG2 | 5.04 | 5 | 1.49 |

2. Supplementary Figures

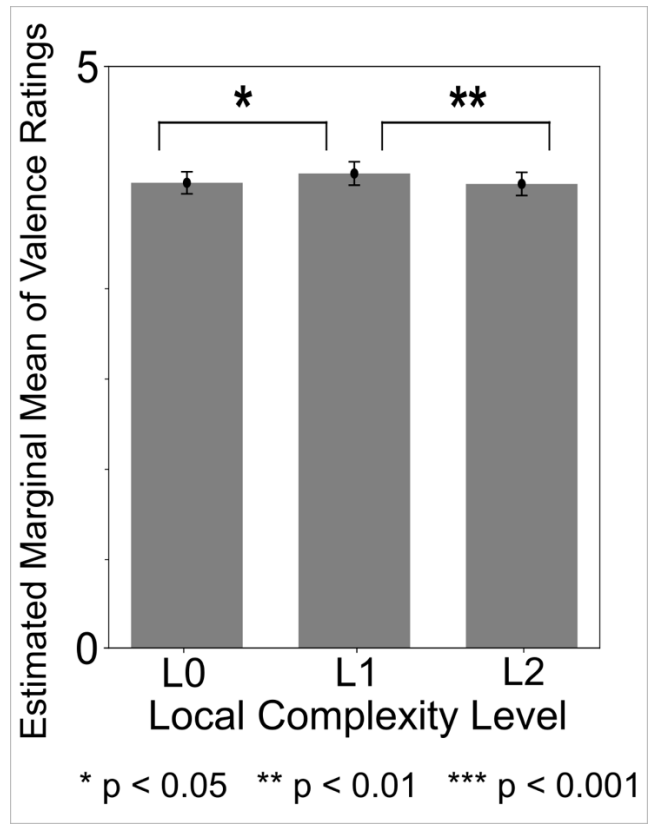

**Supplementary Figure 1. Valence Ratings by Local Complexity Level.** Valence ratings of sound sequences with Local Complexity Level of L1 (medium) were significantly higher than those of L2 (high), regardless of participants' sensory processing patterns or bodily sensations. In some cases, including the Sensory Sensitivity and Upper Torso combination shown in this figure, valence ratings for L1 (medium) were also significantly higher than those for L0 (low).

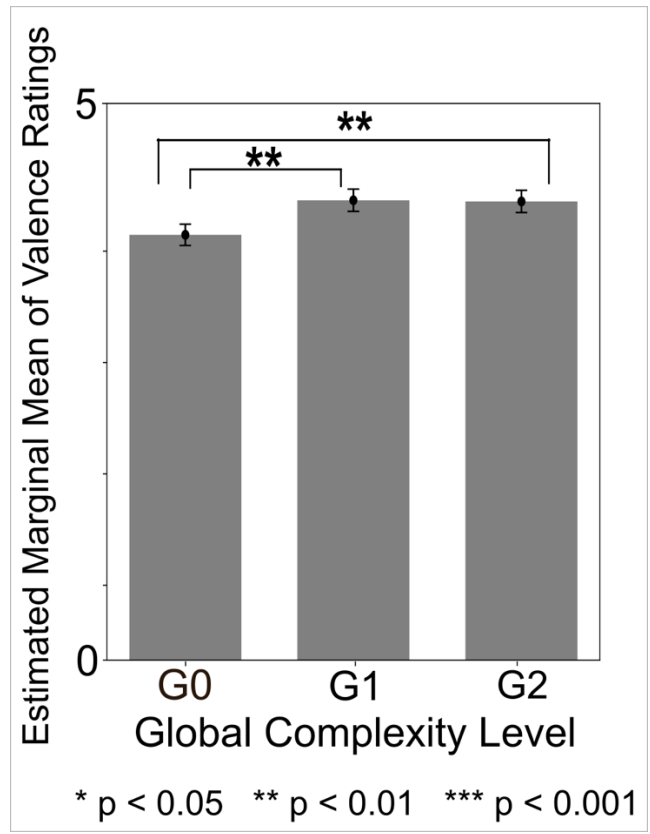

**Supplementary Figure 2. Valence Ratings by Global Complexity Level.** Valence ratings of sound sequences with Global Complexity Level of L0 (low complexity) were lower than those of both G1 and G2 (medium and high complexity), regardless of participants' sensory processing patterns or bodily sensations.

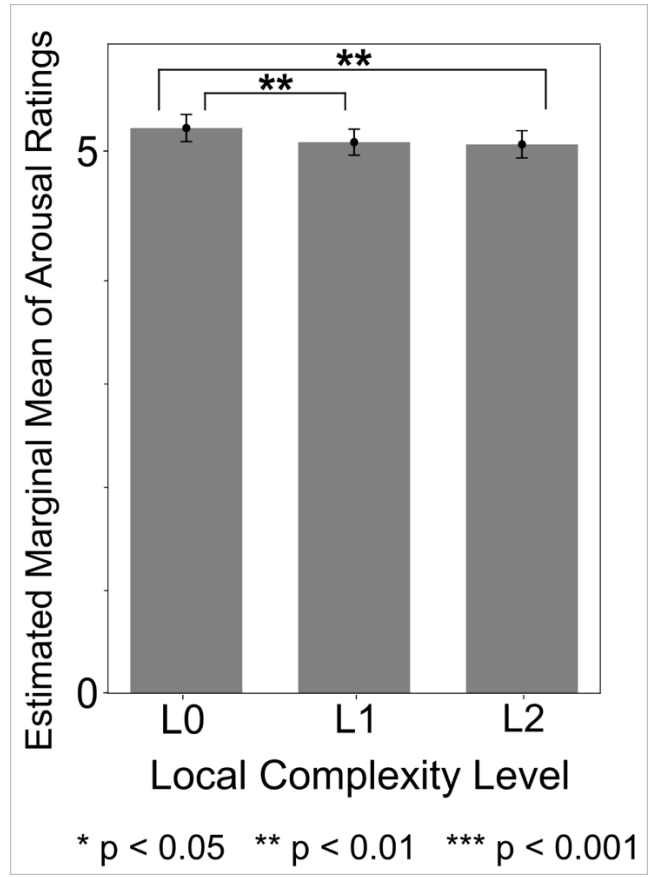

**Supplementary Figure 3. Arousal Ratings by Local Complexity Level.** Arousal ratings were highest for L0 sound sequences (lowest Local Complexity Level), regardless of participants' sensory processing patterns or bodily sensations.

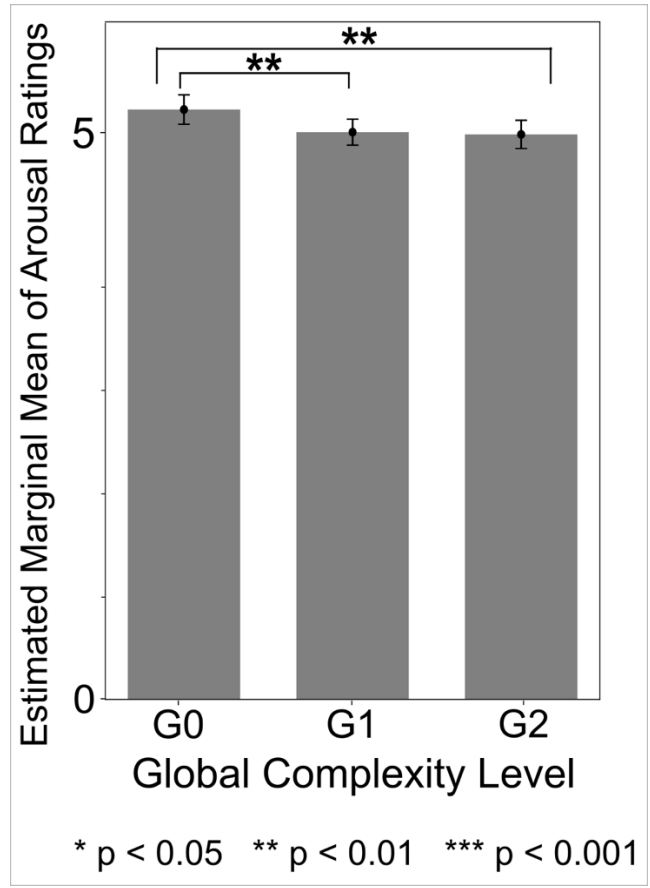

**Supplementary Figure 4. Arousal Ratings by Global Complexity Level.** Arousal ratings were highest for G0 sound sequences (lowest Global Complexity Level), regardless of participants' sensory processing patterns or bodily sensations.

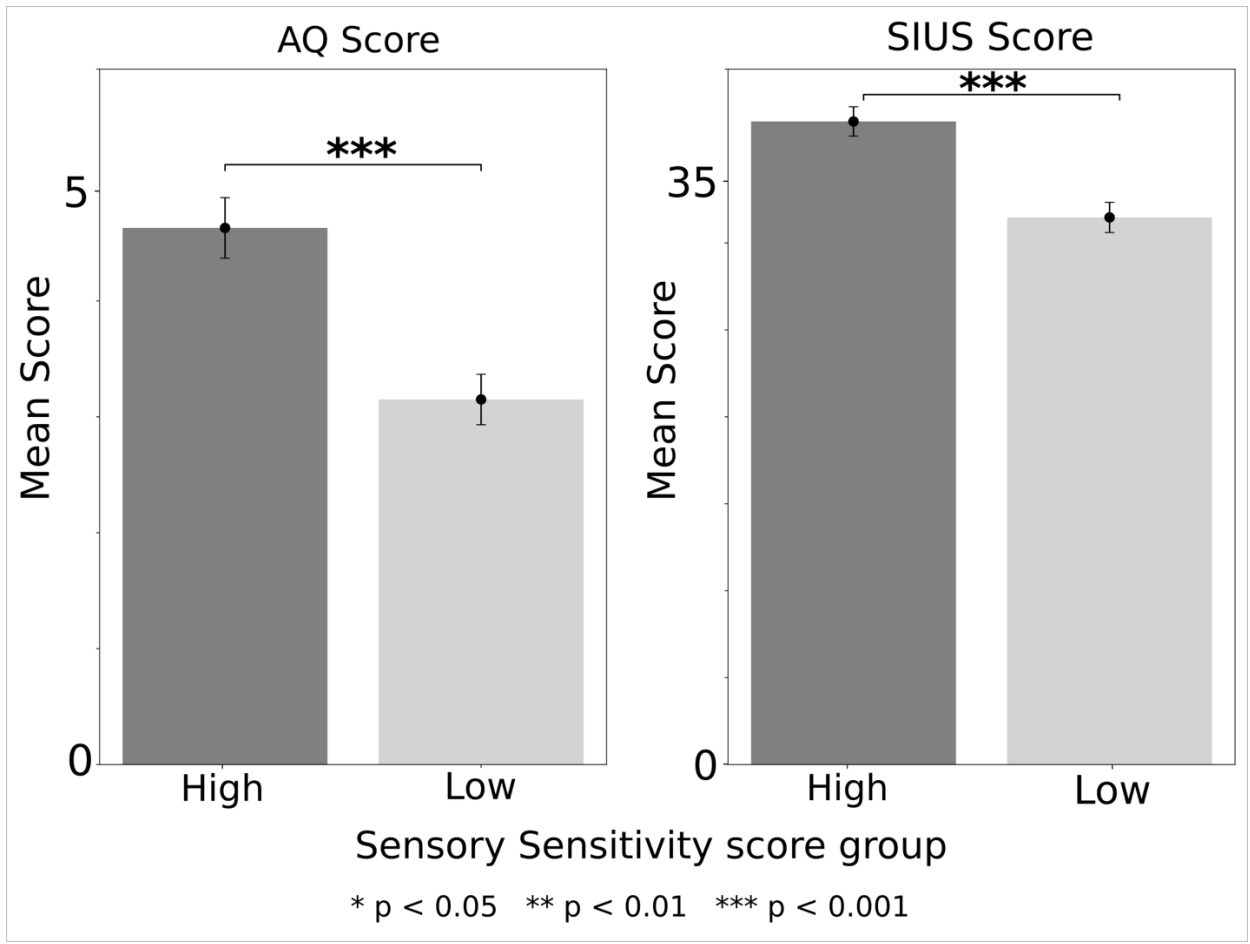

**Supplementary Figure 5. Differences in AQ and SIUS Mean Scores by Sensory Sensitivity Score Group.** Participants with high Sensory Sensitivity scores had significantly higher AQ and SIUS scores than those with low Sensory Sensitivity scores.

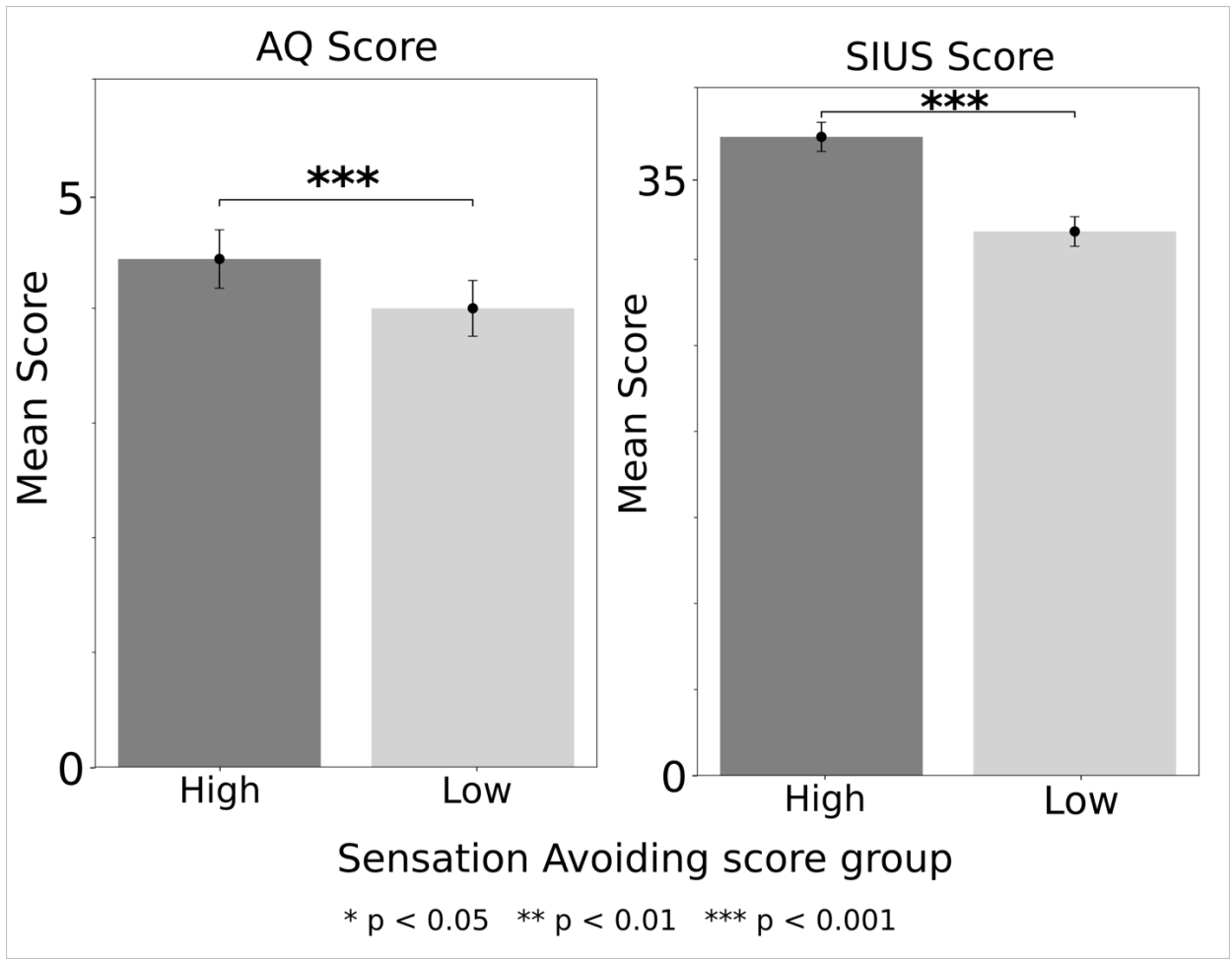

**Supplementary Figure 6. Differences in AQ and SIUS Mean Scores by Sensation Avoiding Score Group.** Participants with high Sensation Avoiding scores had significantly higher AQ and SIUS scores than those with low Sensation Avoiding scores.

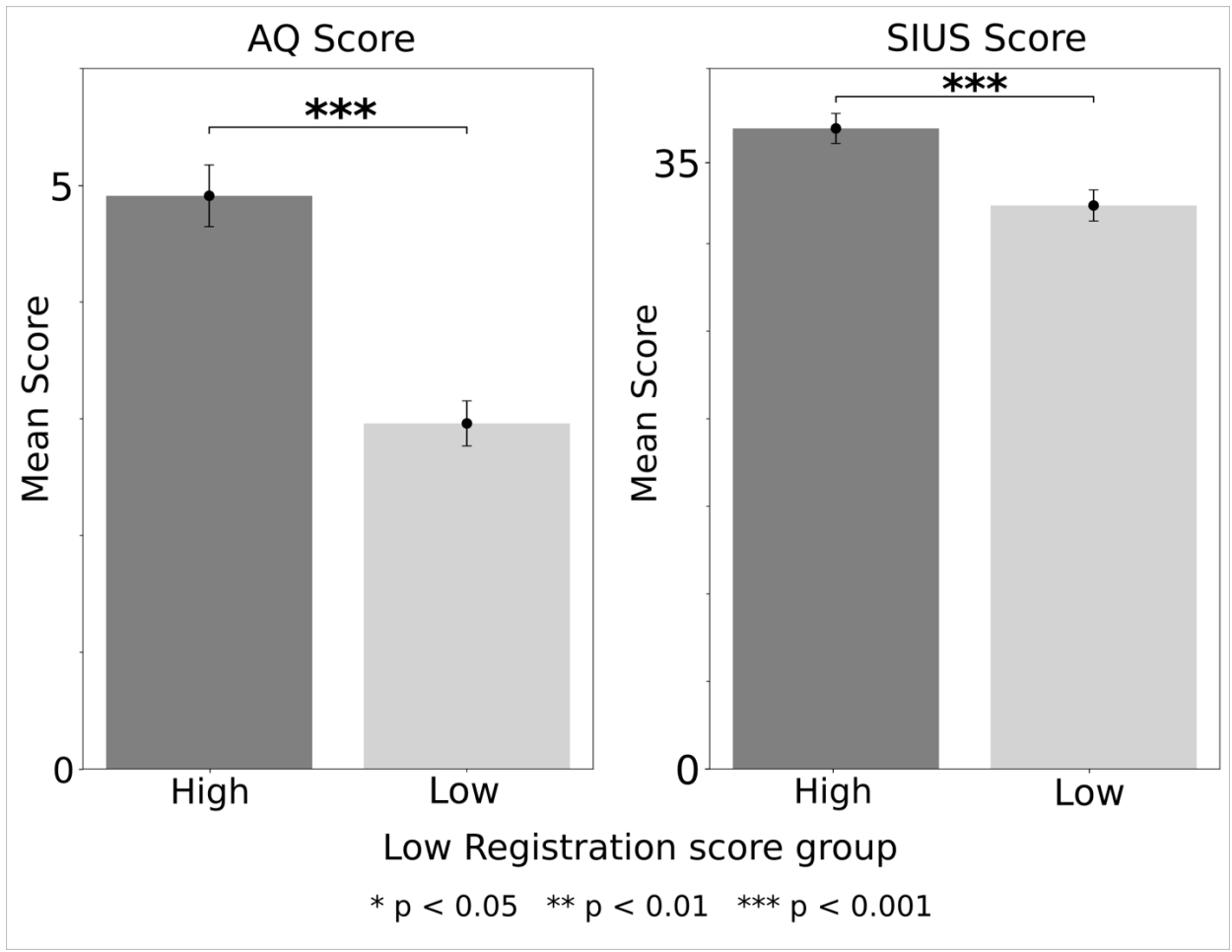

**Supplementary Figure 7. Differences in AQ and SIUS Mean Scores by Low Registration Score Group.** Participants with high Low Registration scores had significantly higher AQ and SIUS scores than those with low Low Registration scores.

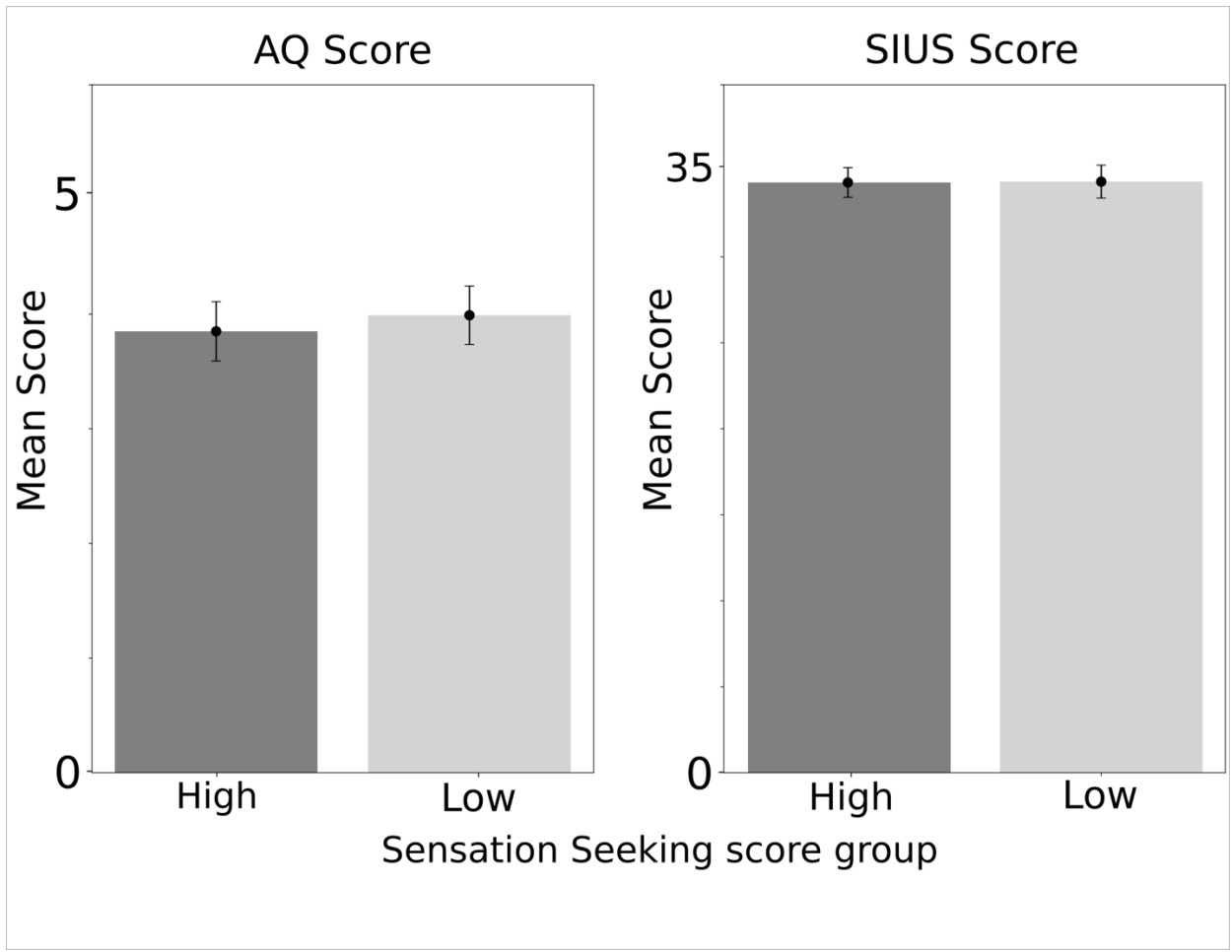

**Supplementary Figure 8. Differences in AQ and SIUS Mean Scores by Sensation Seeking Score Group.** No significant differences were found between AQ or SIUS scores for participants with high versus low Sensation Seeking scores.